\documentclass[twocolumn,pre,amsmath,amssymb,superscriptaddress]{revtex4-2}

\usepackage{graphicx,graphics}
\usepackage{dcolumn}
\usepackage{bm}
\usepackage[normalem]{ulem}
\usepackage[usenames]{color} 
\usepackage{float}

\newcommand{\be}{\begin{equation}}
\newcommand{\ee}{\end{equation}}
\newcommand{\bea}{\begin{eqnarray}}
\newcommand{\eea}{\end{eqnarray}}

\newcommand{\Eb}{E_\text{b}}

\newcommand{\CS}{\text{C}_{60}}

\usepackage[normalem]{ulem}

\begin{document}

\title{
Non-monotonic roughness evolution in film growth on weakly interacting substrates}

\author{Dmitry Lapkin}
\thanks{These authors contributed equally}
\affiliation{Eberhard Karls Universit{\"at} T{\"u}bingen, Institut f{\"u}r Angewandte Physik, Auf der Morgenstelle 10, 72076 T{\"u}bingen, Germany}

\author{Ismael S. S. Carrasco}
\thanks{These authors contributed equally}
\affiliation{International Center of Physics, Institute of Physics, University of Bras\'{\i}lia, 70910-900 Bras\'{\i}lia, Federal District, Brazil}
\affiliation{Departamento de F\'isica, Universidade Federal de Vi\c cosa, 36570-900 Vi\c cosa, MG, Brazil}

\author{Catherine Cruz Luukkonen}
\affiliation{Eberhard Karls Universit{\"at} T{\"u}bingen, Institut f{\"u}r Angewandte Physik, Auf der Morgenstelle 10, 72076 T{\"u}bingen, Germany}

\author{Oleg Konovalov}
\affiliation{European Synchrotron Radiation Facility (ESRF), 71 avenue des Martyrs, 38000 Grenoble, France}

\author{Alexander Hinderhofer}
\affiliation{Eberhard Karls Universit{\"at} T{\"u}bingen, Institut f{\"u}r Angewandte Physik, Auf der Morgenstelle 10, 72076 T{\"u}bingen, Germany}

\author{Frank Schreiber}
\email[Contact author: ]{frank.schreiber@uni-tuebingen.de}
\affiliation{Eberhard Karls Universit{\"at} T{\"u}bingen, Institut f{\"u}r Angewandte Physik, Auf der Morgenstelle 10, 72076 T{\"u}bingen, Germany}

\author{F\'abio D. A. Aar{\~{a}}o Reis}
\email[Contact author: ]{fdaar@protonmail.com}
\affiliation{Instituto de F\'{\i}sica, Universidade Federal Fluminense, Avenida Litor\^{a}nea s/n, 24210-340 Niter\'{o}i, RJ, Brazil}

\author{Martin Oettel}
\email[Contact author: ]{martin.oettel@uni-tuebingen.de}
\affiliation{Eberhard Karls Universit{\"at} T{\"u}bingen, Institut f{\"u}r Angewandte Physik, Auf der Morgenstelle 10, 72076 T{\"u}bingen, Germany}

\date{\today}

\begin{abstract}
Thin film deposition on weakly interacting substrates exhibits a unique growth mode characterized by initially strong island formation and rapidly increasing roughness, which reaches a maximum and subsequently decreases as the film returns to a smooth morphology.
Here we show this rough-to-smooth growth mode experimentally for two molecular systems with substantially different geometries, namely, the effectively spherical buckminsterfullerene ($\CS$)  and the disk-like 1,4,5,8,9,11-hexaazatriphenylenehexacarbonitrile (HATCN).
This growth mode is explained by a geometrical model that captures the basic mechanisms of multilayer island growth, island coalescence, and formation of a continuous film. Additionally, kinetic Monte Carlo simulations with minimal ingredients demonstrate that this mode generally occurs for weakly interacting substrates, providing quantitative estimates of parameters that characterize adsorbate-adsorbate and adsorbate-substrate interactions.
Both the model and simulations accurately describe the experimental data and highlight the generic nature of the phenomenon, independently of the details of the interactions and the molecular flux, which opens up a path for controlling nanoscale film roughness.
\end{abstract}

\maketitle

Thin film growth is one of the most important processes in modern science and technology \citep{ohring}, but a complete understanding of its basic nonequilibrium processes is yet to be found.
Combined insights from experiments and theoretical approaches were very fruitful in explaining selected scenarios of both the early and late stages of metal and semiconductor growth on their own surfaces (A-on-A growth) as a self-organization process with atom deposition and surface diffusion as basic microscopic dynamic ingredients \cite{venables,pimpinelli,ratsch2003,krug,michely,etb,einax}.
However, A-on-B growth saw much less progress in terms of theoretical understanding and has remained a long-standing problem, despite being the technologically more relevant case.
The regimes of island formation and of smooth growth (layer-by-layer [LBL]) of A-on-B growth have been frequently related to the near-equilibrium regimes of partial and full wetting of material A on substrate B, respectively \cite{venables1984_review,brune2001,burkeJPCM2009}, whereas far-from-equilibrium approaches represent the film growth phenomenon as an unbinding transition, e.g.~in certain lattice models or continuum treatments such as the Kardar-Parisi-Zhang equation \cite{hinrichsen1997,munoz1998,giada2000}.
Such concepts, although being of great general value, are insufficient to describe the experimentally observed features of A-on-B growth, particularly in the earliest stages of deposition. 
In terms of morphology, the initial formation of multilayer islands is frequently observed \citep{elofsson2014,liuMatResExp2017,parveen2020,toApplSS2021,Reisz_2021}, particularly if B is a weakly interacting substrate.
A nontrivial outcome observed experimentally \citep{liuMatResExp2017,parveen2020,toApplSS2021,Reisz_2021,drakopoulou2023} and in simulations \citep{toApplSS2021,empting2021,toreis2022} is a non-monotonic dependence of the roughness on the thickness, which is in stark contrast to the monotonic roughness increase with time in theories of kinetic roughening of interfaces, often formulated as scaling laws \cite{kpz,ramasco,durr2003_prl}.

In this letter, we report and explain the generic phenomenon of non-monotonic roughness evolution in A-on-B growth with a combination of experimental studies on organic films, theoretical methods, and kinetic Monte Carlo (KMC) simulations.
The {\it rough-to-smooth growth mode} is characterized by initial multilayer island formation and rapidly growing roughness, followed by island coalescence after a maximal roughness is attained, and return to smooth film morphologies with a roughness of 1\textendash 2 unit cell lengths.
The conceptual explanation of this mode is provided by a geometrical model, and KMC simulations show that the mode occurs for broad ranges of diffusion-to-deposition ratios and activation energies of surface diffusion coefficients, which supports its generic nature. The microscopic physical mechanism leading to this growth mode is the enhanced hopping rate of adsorbates
from the substrate to the film.
This picture differs from near-equilibrium approaches (e.g.~Volmer-Weber mode \citep{ohring} and partial wetting models \citep{binder1992,empting2021}) and goes beyond A-on-A growth theories due to the explicit influence of substrate B in the early stages of growth.
Due to the ubiquitous nature of thin films in technological applications, the A-on-B case is highly relevant and thin films with very low roughness are indeed desirable for many devices \citep{klipfel2022,dongOE2019}.

\begin{figure}[!h]
\includegraphics[width=\columnwidth]{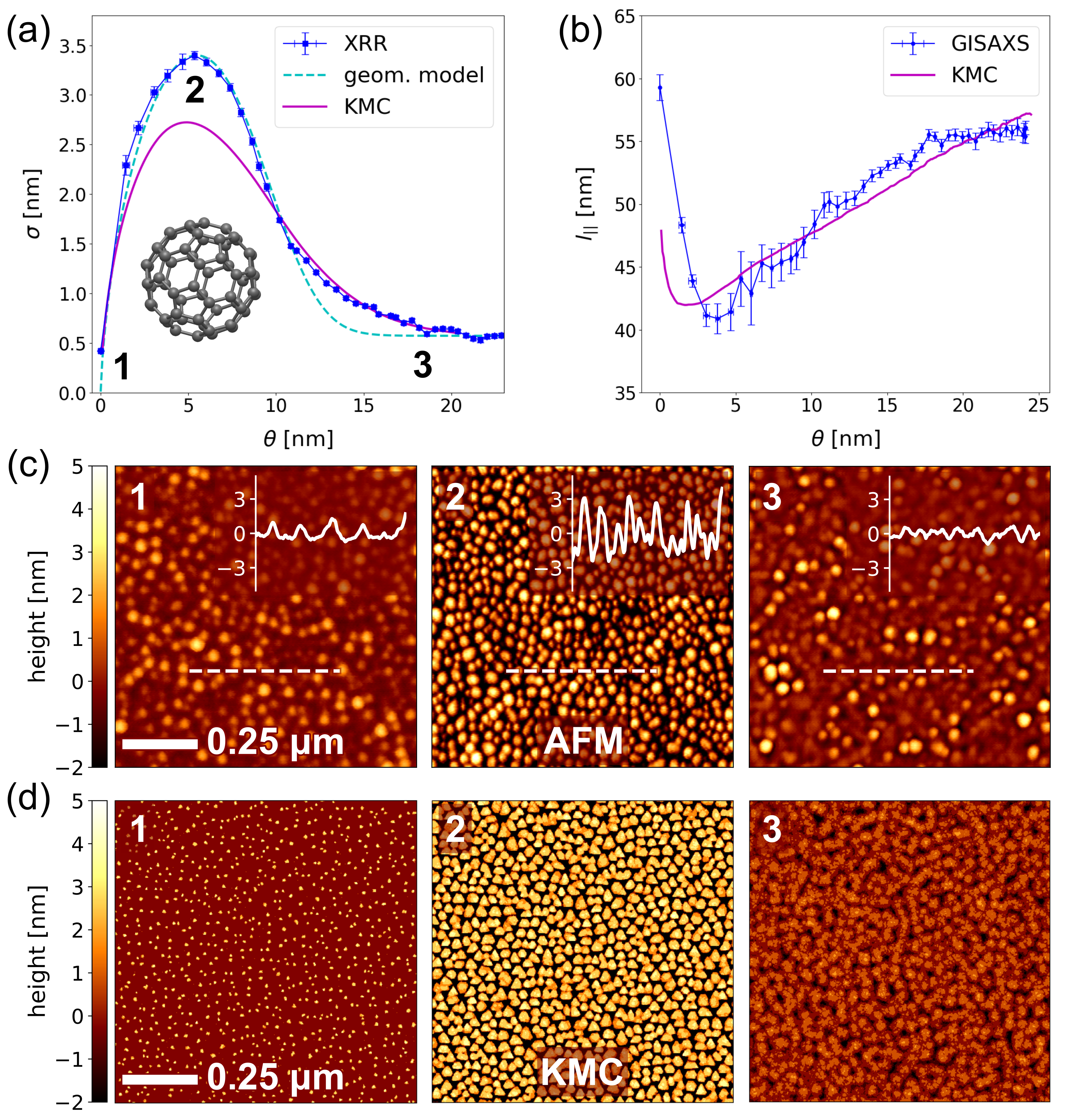}
\caption{(Color online) (a) Roughness $\sigma$ and (b) in-plane correlation length $l_{\parallel}$ evolutions of $\CS$ films.
Thin full lines (blue; darker gray) with error bars are the experimentally extracted dependencies.
Thick full lines (magenta; lighter gray) are KMC simulation results with $R=3\cdot10^3$, $E_b=2.5$, $E_s/E_b=1.33$.
The dashed line in (a) is a fit to the geometric model (parameters $\alpha=0.28$, $C_L=5\%$, $L/d_0=2.0$, and $\beta d_0=0.115$). The inset in (a) shows the chemical structure of the $\CS$ molecule with carbon atoms represented as gray spheres.
(c) AFM images of $\CS$ film at different thicknesses indicated in (a): (\textbf{1}) start of the growth; (\textbf{2}) roughness maximum; (\textbf{3}) smooth growth. The insets show exemplary line profiles through the maps taken at the positions indicated with white dashed lines. The horizontal axes are identical to those of the maps.
(d) Heatmap height profiles from the KMC simulations under the same conditions.
}
\label{fig:roughness}
\end{figure}

We study thin films with different molecular shapes relevant for organic electronics and grown on silicon substrates with a native silicon oxide layer (Si/SiO$_x$): buckminsterfullerene ($\CS$) with essentially spherical molecules (diameter $1.0$~nm, the chemical structure is shown in Fig.~\ref{fig:roughness}(a)) and 1,4,5,8,9,11-hexaazatriphenylenehexacarbonitrile (HATCN) with flat disk-like shape (diameter $1\text{--}1.5$~nm, height $0.3$~nm).
$\CS$ is an n-type organic semiconductor with high charge carrier mobility and good thermal, chemical, and UV-light stability in its crystalline form.
Thin $\CS$ films are used in organic field effect transistors \citep{kyndiah2015,huttner2019,qiaoACSANM2023}, solar cells \citep{klipfel2022,artuk2024}, and phototransistors \cite{choiACSAMI2023}.
Conversely, HATCN is a strongly electron-deficient molecule suitable for hole injection in organic light-emitting diodes (OLEDs) and organic photovoltaic devices \citep{kimJMChem2013,winkler2016,leeJMCC2017,juangICCCPhot2022,yazdani2022,liangAM2023}.
Thin $2\text{--}10$~nm HATCN films with uniform thickness increase the efficiency and reduce thermal damage effects in quantum dot LEDs \citep{dongOE2019,liuACSAMI2021}.

Thin films of $\CS$ were deposited on Si/SiO$_x$ substrates ($10\times52$~mm$^2$) using a specially designed organic molecular beam deposition chamber with a moving mask \cite{Lapkin2025}.
The samples had a linear thickness gradient along the long substrate axis, where the film thickness varied from 0 to $\sim20$~nm.
A non-invasive study of the film morphology was performed with spatially-resolved X-ray scattering techniques: X-ray reflectometry (XRR) with a General Electric XRD~3003~TT diffractometer and grazing-incidence small-angle X-ray scattering (GISAXS) performed at the ID10-SURF beamline at the ESRF.
The measurements were performed every $1$~mm along the thickness gradient (larger than the X-ray beam size) and provide the roughness and in-plane correlation length as functions of the thickness shown in Figs.~\ref{fig:roughness}(a) and \ref{fig:roughness}(b), respectively
(see Sec.~S.I of the Supplemental Material (SM)~\cite{supplementary} for details of experiments and data analysis.).
Complementary real space information was obtained by Atomic Force Microscopy (AFM) (JPK NanoWizard~II); see Fig.~\ref{fig:roughness}(c).
See Sec.~S.III of the SM~\cite{supplementary} for reproducibility beyond the data shown in Fig.~\ref{fig:roughness}.
The $\CS$ films are polycrystalline with randomly oriented face-centered cubic (fcc) grains with a lattice constant of $\sim 1.42$~nm \cite{heiney1991}; accordingly, the distance between (111) layers $\sim 0.82$~nm corresponds to a monolayer (ML) unit of this material.
HATCN films deposited and analyzed by the same techniques show a polycrystalline structure with a low degree of orientation and a large trigonal unit cell \cite{yadav2020}.

The $\CS$ films show the rough-to-smooth growth mode: the roughness $\sigma$ rapidly increases until a maximum of $\sigma\approx3.5$~nm at a thickness $\theta\approx5.4$~nm and slightly slower decreases towards a saturating roughness  $\sigma<1$~nm, which is reached at a thickness $\theta\approx20$~nm.
For submonolayer coverages, the AFM image \textbf{1} in Fig.~\ref{fig:roughness}(c) shows isolated, multilayer islands;
at the point of maximum roughness (\textbf{2} in Fig.~\ref{fig:roughness}(c)), those islands have grown further and have nearly coalesced; close to the roughness saturation (\textbf{3} in Fig.~\ref{fig:roughness}(c)) the islands have fully coalesced and the inter-island trenches have been largely filled.
In Fig.~\ref{fig:roughness}(b), the initial in-plane correlation length of $l_{||}\approx60$~nm is caused by sub-nm inhomogeneities of the bare substrate with the roughness $\sigma_\text{sub}\approx0.4$~nm (see Sec.~S.II in the SM~\cite{supplementary} for details).
The in-plane correlation length $l_{||}$ approximately corresponds to the inter-island distance (Fig.~\ref{fig:roughness}(c)). 
$l_{||}$ decreases while the islands grow until a thickness of $\theta\approx4$~nm is reached.
Subsequently, $l_{||}$ slowly increases and the island density $n_\text{isl} \sim 1/l_{||}^2$ decreases, reflecting island coalescence.  
Additional AFM data are provided in Secs.~S.I--S.III of the SM~\cite{supplementary}.

\begin{figure}[!h]
\includegraphics[width=\columnwidth]{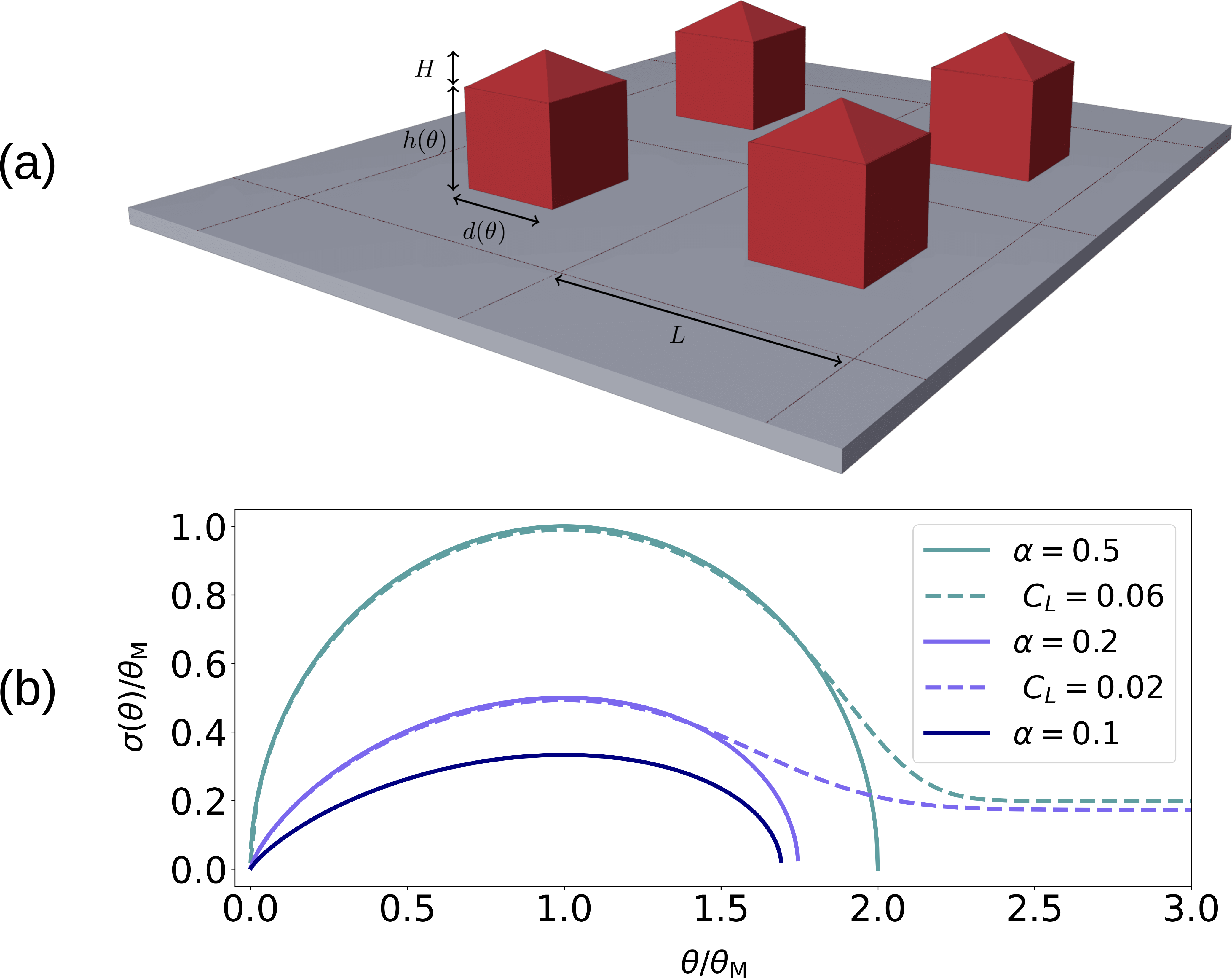}
\caption{(a) Geometric model with islands growing with power laws for $d(\theta)$ and $h(\theta)$ in a capture zone of size $L \times L$.
(b) Reduced roughness $\sigma/\theta_M$ vs. $\theta/\theta_M$ from Eq.~\ref{sigmascaling} (full lines) and from the full geometric model (dashed lines) with $C_L=0.02$ for $\alpha=0.2$, $C_L=0.06$ for $\alpha=0.5$, and a reduced slope parameter for the pyramidal cover $\beta d_0=0.2$.}
\label{fig:toymodel}
\end{figure} 


Previous experiments and modeling \citep{etb,PE2007,einax} of submonolayer growth show that island densities typically saturate at small coverages below one ML, and that each island subsequently grows by aggregation of diffusing atoms deposited in its capture zone.
Based on these findings, we propose the geometric model for multilayer island growth illustrated in Fig. \ref{fig:toymodel}(a): each island grows in a capture zone whose width $L$ obeys a Gaussian distribution with relative standard deviation $C_L$; the islands have the shape of rectangular parallelepipeds (width $d$, height $h$) with pyramidal covers (height $H$) of constant slope; these characteristic sizes increase as power laws of the thickness $\theta$ until $d=L$ (when the capture zone is fully covered) and, subsequently, only the height $h$ increases.
We write $d=d_0\theta^\alpha$ and $H=3 \beta d_0\theta^{\alpha}$ (where $\beta$ and $d_0$ are constants); this pyramidal cover mimics a mound with selected slope, which is a feature related to ES barriers \cite{siegertPRE1996,amarPRB1999,luisMTC2025}. 
All lengths are measured here in ML units.
These relations entail $h= L^2/d_0^2 \theta^{1-2\alpha}-\beta d_0 \theta^\alpha$, which shows that the exponent $0<\alpha<0.5$ controls the aspect ratio evolution, with larger $\alpha$ for broader islands.

The simplest version of this model has $\beta=0$ (flat island covers) and $C_L=0$ (all capture zones have the same width).
As $\theta$ increases, the roughness $\sigma$ increases, reaches a maximum $\sigma_M$ at thickness $\theta_M$, and then decreases to zero at a thickness $\theta_c$ in which the islands coalesce ($d(\theta_c)=L$).
We obtain $\theta_M={\left( 1-\alpha\right)}^{1/\left( 2\alpha\right)}{\left( L/d_0\right)}^{1/\alpha}$ and $\theta_c={\left( L/d_0\right)}^{1/\alpha}$; the former plays the role of a scaling factor, so that the roughness evolution and the maximal roughness are given in scaled forms as
\be
\frac{\sigma}{\theta_M}=\frac{\theta}{\theta_M} \sqrt{ \frac{1}{ 1-\alpha} {\left( \frac{\theta}{\theta_M}\right)}^{-2\alpha} -1 } \ , \quad \frac{\sigma_M}{\theta_M}=\sqrt{\frac{\alpha}{1-\alpha}} \; .
\label{sigmascaling}
\ee
The roughness versus thickness curve has a negative skewness whose absolute value decreases as $\alpha$ increases, as illustrated in Fig. \ref{fig:toymodel}(b), and becomes symmetric for $\alpha=1/2$.
This captures the main features of the rough-to-smooth growth mode of $\CS$ films: using the experimental estimate $\sigma_M/\theta_M \approx 0.6$ (Fig.~\ref{fig:roughness}(a)), the model yields $\alpha\approx0.25$ and provides a reasonable fit of the roughness peak, as shown in Fig. S10 of the SM~\cite{supplementary}.

This geometric description of rough-to-smooth growth mode is improved when we account for $C_L>0$  (see Sec.~S.V of the SM~\cite{supplementary} for details)  and for the pyramidal cover ($\beta>0$).
Figure \ref{fig:toymodel}(b) shows that even a small value of $C_L$ can lead to slower decay of the roughness after the peak and that a non-zero asymptotic roughness is set by the pyramidal cover.
Fig. \ref{fig:roughness}(a) shows a very good fit of the $\CS$ film data by this full geometric model, the low value of the saturation roughness at the largest widths and the small value of $\beta$ used in this fit suggest a negligible ES barrier.

\begin{figure}[!h]
\includegraphics[width=\columnwidth]{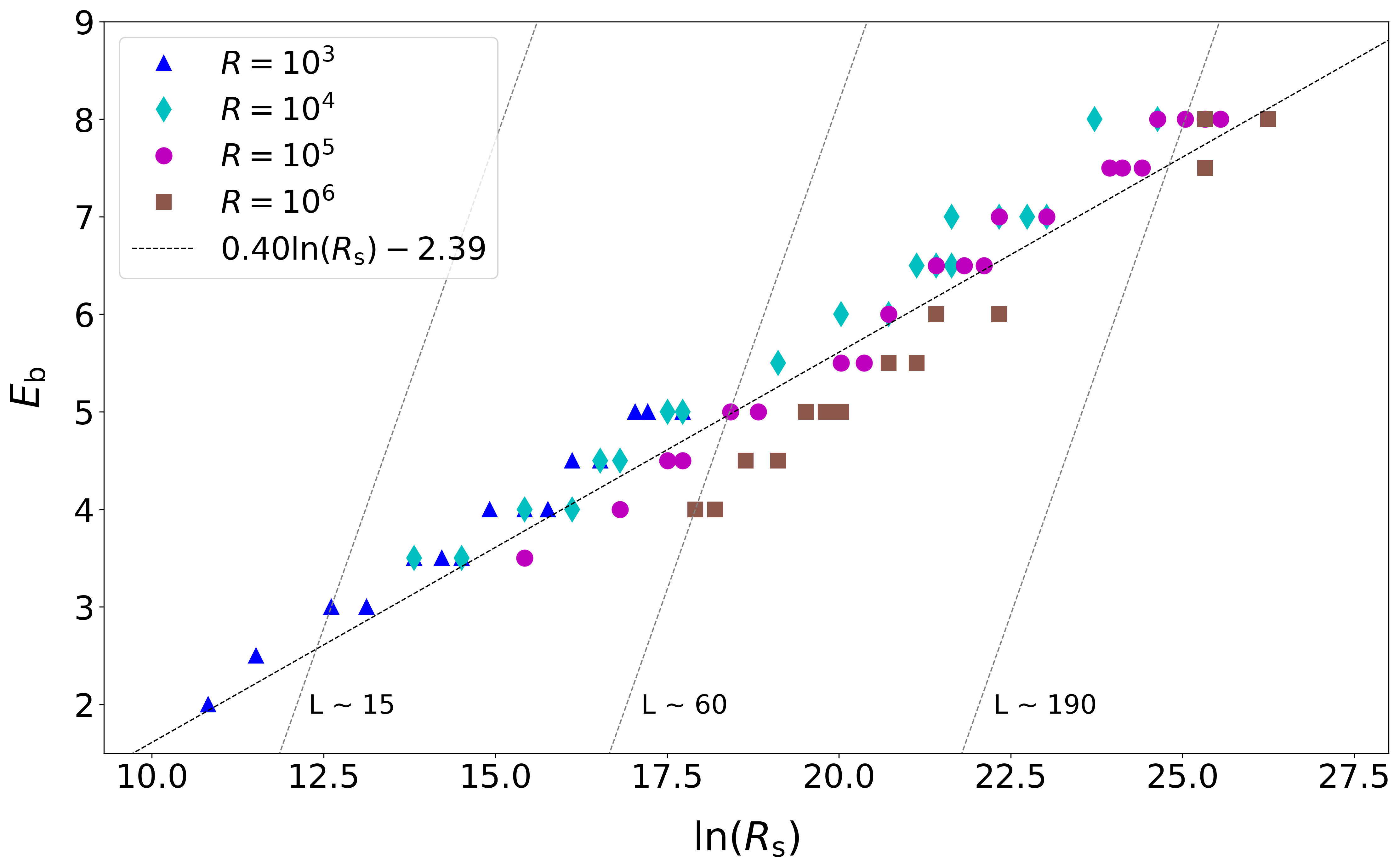}
\caption{KMC parameters $\{R,R_s,E_b\}$ from the set $\mathcal{S}$
(roughness maximum at coverages $5\leq\theta_M\leq 8$) fall onto an approximate line in the  $\Eb-\ln R_s$ plane. The different colored symbols differ in their $R$ value for the diffusion of film particles on film. 
The dotted lines are lines of constant $L=R_s^{1/4}\exp(-E_b/8)$, which is the approximate capture zone size (island distance) in ML units
according to submonolayer scaling theory \cite{oliveirareis2013}.}
\label{Eb-RS-diagram}
\end{figure}


A frequently observed feature in organic film deposition is layer-by-layer growth (oscillating roughness $<1$~ML) followed by rapid roughening in some form of unstable growth \citep{kowarik2007,kowarikJPCM2008,frankPRB2014a,lorchAPL2015,chiodini2020}.
This transition can be quantitatively described with KMC simulations of A-on-A growth models with ES barriers \citep{kang1992,smilauerPRB1995,bartelt1999,luisMTC2025}.
On the other hand, the formation of islands on the bare substrate in A-on-B growth has been addressed in KMC studies using larger adatom diffusion coefficients on the substrate ($D_S$) than those on the upper layers of the film ($D$) and using no ES barrier \citep{toApplSS2021,empting2021,toreis2022}. Building on that, we have defined a Clarke-Vvedensky-type deposition model \citep{CV,etb} on an fcc lattice (the same as crystalline films of $\CS$) where each molecule occupies a lattice site of edge $a$, the substrate (layer $0$) is a (111) plane, the vertical flux is $F$ (number of incident molecules per substrate site per unit time), and the hopping rates of molecules on the film (layers $\geq2$) are $D_n=\nu\exp{\left(-E_D-n E_b\right)}$, where $n$ is the number of lateral nearest neighbors (NN) with binding energy $E_b$ each, $E_D$ corresponds to the interaction with the layer below, and $\nu$ is a hopping attempt frequency. All energies are in units of the Boltzmann constant times temperature ($k_B T$).
Considering the relatively large size of the organic molecules and the short range of van der Waals interactions, we use $E_D=3E_b$ corresponding to bonds with three nearest-neighbors in the underlying layer of the fcc structure.
The molecules in layer $1$ have hopping rate $D_{sn}=\nu\exp{\left(-E_s-nE_b\right)}$, where $E_s<3E_b$ accounts for a weaker interaction with the substrate and a correspondingly larger mobility.
The dimensionless ratio $R=D_0/F$ controls the interplay of surface diffusion and flux in the upper layers, whereas $R_s=D_{s0}/F=R\exp(3E_b-E_s)$ plays this role for molecules on the substrate.
Further details of the KMC simulations and of the calculation of physical quantities are provided in Sec.~S.VI of the SM~\cite{supplementary}.

The parameter space $\{R,R_s,E_b\}$ of the model can be considered truly minimal for describing A-on-B growth. The two variables $R$ and $E_b$ are sufficient to describe A-on-A growth \cite{etb,einax,venables,krug} and the single parameter $R_s$ (containing the substrate energy $E_s$) captures the substrate effect.
We recover two well-defined limiting cases: layer-by-layer epitaxial growth ($R_s/R=1$) and near-equilibrium growth ($R \to \infty$) leading to partial wetting of the substrate (Vollmer-Weber growth), see also Ref.~\cite{munko2025}.
Since Fig.~\ref{fig:roughness}(a) shows $\theta_M \approx 6.6$ ML (5.4 nm) with some spread, we performed a search within the parameter space for roughness curves with $5~\text{ML}\leq\theta_M\leq 8~\text{ML}$, which provided us with a set $\mathcal{S}$ of suitable parameters for $\CS$ films.
The scaled roughness curves $\sigma/\theta_M$ have the same principal shape as the curves from the geometric model; see Fig.~S12 in the SM~\cite{supplementary} for a complete representation.
This is indicative of the generic nature of the rough-to-smooth phenomenon: a novel growth mode induced primarily by surface energy differences between substrate and film, but otherwise largely independent of the specific value for the deposition speed (embodied in $R$).
Sec. S.IX of the SM~\cite{supplementary} discusses further the conditions on the magnitude of $E_s$ for the growth mode to appear.
Figure \ref{Eb-RS-diagram} shows that set $\mathcal{S}$ defines an approximate line in the $E_b-\ln R_s$ plane (independent of $R$), in which $\exp(-\Eb)R_s^\gamma$ is approximately constant with exponent $\gamma\approx0.40$.
This line is fully inside the regime of unstable dimers and stable triangles of submonolayer growth theory on a triangular lattice \citep{oliveirareis2013}, which is a condition where stable islands mostly grow compactly (with flat borders), as assumed in the geometric model.

The same theory predicts an island density of order $\sim\exp(\Eb/4)R_s^{-1/2}$ in this regime, so we define an island distance parameter $L=\exp(-\Eb/8)R_S^{1/4}$ (in ML units).
Figure \ref{Eb-RS-diagram} shows some lines of constant $L$, which are much steeper than the line of set $\mathcal{S}$.
The AFM island density at the roughness maximum (point \textbf{2} in Fig.~\ref{fig:roughness}(c)) is obtained in simulations within set $\mathcal{S}$ for $E_b=2.5$ and $E_s=1.33E_b$, which gives $L\approx 16$ (which is the same order of magnitude as the experimental island distance $\approx$ 50 (in ML units)).
The simulational surface height profile maps are shown in Fig.~\ref{fig:roughness}(d) for the points \textbf{1}, \textbf{2}, and \textbf{3}, which correspond to the same thicknesses of the experimental images of Fig.~\ref{fig:roughness}(c).
The KMC roughness curve, shown in Fig.~\ref{fig:roughness}(b), has a peak with the same position and width as the XRR data, but somewhat smaller height.
This can be rationalized by observing that the island height distribution in the AFM images is broad (Fig.~\ref{fig:roughness}(c)), while these heights are rather uniform in the KMC simulations (Fig.~\ref{fig:roughness}(d)).
This is presumably related to the random orientation of crystalline grains in the experimental sample, which is not taken into account in the KMC simulations.
The in-plane correlation length from KMC simulations (island-island distance as determined from the second maximum of the autocorrelation function), shown in Fig.~\ref{fig:roughness}(b), shows a minimum and then increases, in rather good quantitative agreement with the experimental correlation length.

We note that our KMC estimate for $E_b=2.5$ corresponds to $\approx 64$~meV at room temperature, which should be interpreted as an effective value
for the specific case of polycrystalline $\CS$ growth,
as it is $2$ to $4$ times smaller than the literature values for more ordered films
($E_b=130$~meV in \citep{bommel2014} and $E_b\approx235$~meV in \citep{janke2020}).
Note also that our KMC simulations lead to smaller surface diffusion coefficients of free $\CS$ molecules than previous MD simulations \cite{cantrell_clancy2008} (see details in Sec.~S.VIII of the SM~\cite{supplementary}), but this does not affect the overall conclusions on the mechanisms underlying the rough-to-smooth growth mode.

\begin{figure}[!h]
 \includegraphics[width=0.8\columnwidth]{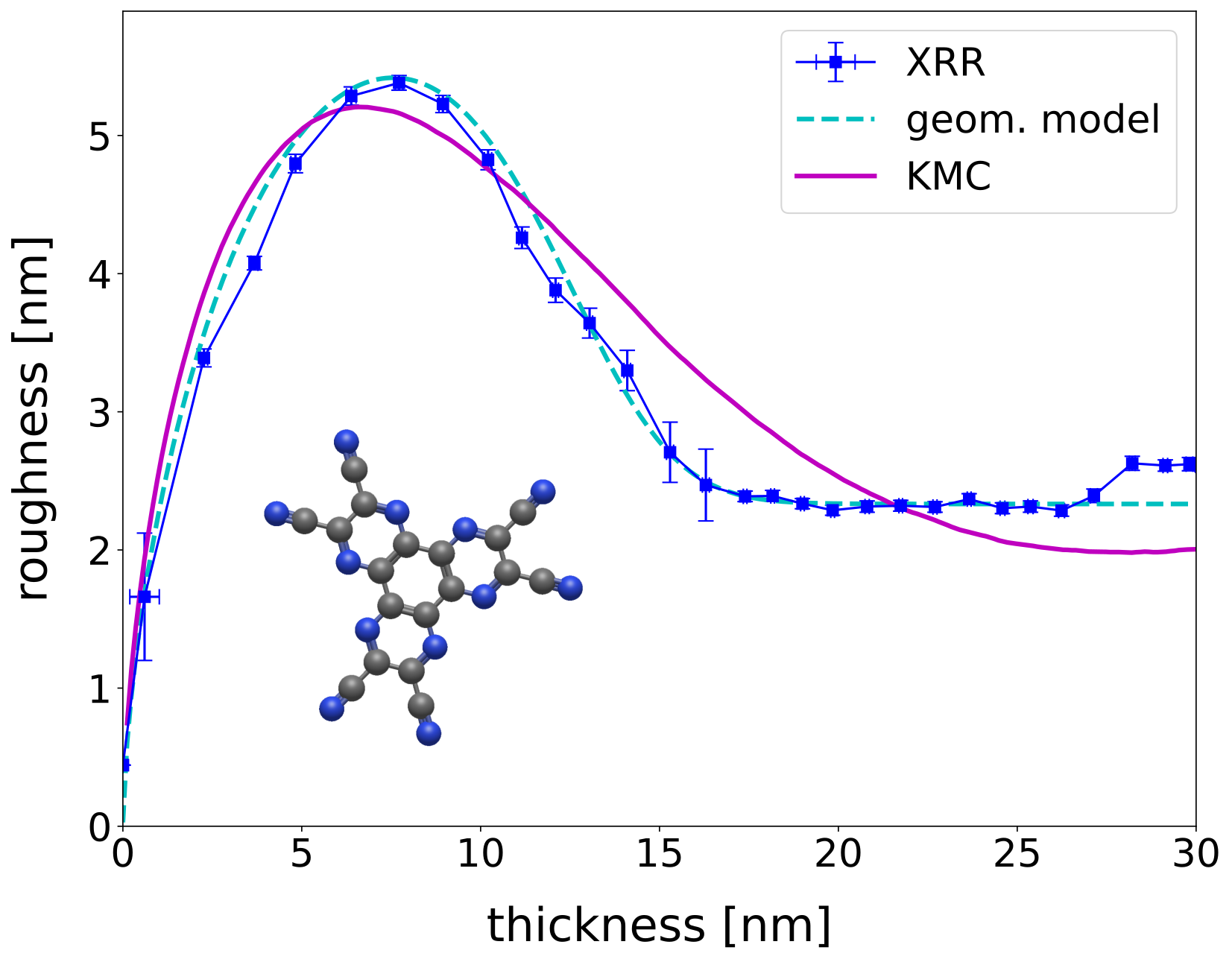}
\caption{Roughness $\sigma$ evolution for HATCN as a function of film thickness $\theta$. Thin full line
(blue) with error bars is the experimentally
extracted (from XRR) dependency.
The dashed line is a fit to the geometric model with Gaussian distributed capture zone areas ($\alpha=0.35$, $L/d_0=2.48$, $C_L=5\%$, $\beta d_0=0.343$).
The KMC parameters for the modified model (see S.VI. of the SM~\cite{supplementary}) are $R=10^5$, $E_b=5.0$, $E_s/E_b = 0.38$ and $E_{ES}=2.3$. The aspect ratio of the tetragonal unit cell is 1.58.
The inset shows the chemical structure of the HATCN molecule with carbon atoms represented as gray spheres and nitrogen atoms as blue spheres.
}
\label{fig:HATCNroughness}
\end{figure}

Figure \ref{fig:HATCNroughness} shows a similar rough-to-smooth
growth mode for HATCN films deposited on Si/SiO$_{x}$ substrates, which is further confirmed by AFM measurements shown in Sec.~S.XI of the SM~\cite{supplementary}. However, $\sigma_M/\theta_M\approx0.73$ is somewhat larger than that ratio for $\CS$ and, after decaying, the roughness exceeds $2$~nm for the thickest films.
The full roughness curve is very well described with the geometric model (Fig.~\ref{fig:HATCNroughness}) with exponent $\alpha=0.35$; compared to $\alpha=0.28$ for $\CS$, this implies a faster increase of the island widths.
The most important difference, though, is the large value of $\beta d_0$, which means that the pyramidal covers of the islands have non-negligible heights and suggests the presence of non-negligible ES barriers to build such mounded structures.

The HATCN molecule, with its disk-like shape (the chemical
structure is shown in Fig.~\ref{fig:HATCNroughness}) as well as its trigonal crystal lattice, is markedly different from the near-spherical $\CS$ molecule with an fcc crystal lattice.
To account for that, we have extended our minimal KMC model by employing a tetragonal lattice with a larger lattice constant in direction perpendicular to the substrate and with different microscopic motion for upright standing molecules, see Sec.~S.VI. of the SM~\cite{supplementary} for more details. The resulting roughness curve (thick full line in Fig.~\ref{fig:HATCNroughness}) fits the peak rather well. 
The decay to a higher roughness compared to the $\CS$ case requires the inclusion of an ES barrier with small energy $E_{ES}=2.3$, which is consistent with the prediction of the geometric model.
For higher ES barriers, the roughness peak disappears altogether and the roughness increases monotonically; the effects of an ES barrier on the KMC results are further discussed in Sec.~S.X of the SM~\cite{supplementary}.

In summary, we have presented novel physical insight for one of the key challenges in thin film deposition, namely A-on-B as opposed to A-on-A growth.
The striking rough-to-smooth growth mode has been established through a quantitative match between experiment, an intuitive geometric model, and microscopic KMC simulations of $\CS$ and HATCN films on weakly interacting substrates.
The simulations and theoretical arguments indicate that the initial roughening and subsequent reversal to a nearly flat film is a rather universal, surface-energy-induced mode, which requires a sufficiently weakly interacting substrate and weak Ehrlich-Schwoebel barriers, but holds for wide ranges of interaction energies and molecular fluxes.
This sheds new light on the connection between growth modes and thermodynamic wetting transitions.  Recently, it has been theoretically shown \cite{munko2025} that LBL growth still holds for moderately weak substrates in the near-equilibrium partial wetting regime and the onset of island growth only occurs for even weaker substrates.
However, here we show that this initial island growth is transitory and that the films recover smooth growth.
The physical mechanism responsible for this non-equilibrium multilayer island growth is the enhanced hopping rate of adsorbates from the weakly interacting substrate to the film \citep{munko2025}, but this driving force vanishes once the islands coalesce and a uniform film with a lower roughness can be formed.
The generic nature of this mode is supported by the similar roughness evolution in $\CS$ and HATCN films on the same substrate, despite the large differences in shape and crystal structure between those molecules, as well as by previous illustrations in inorganic materials \citep{liuMatResExp2017,parveen2020,toApplSS2021}.
The theoretical framework introduced here can quantitatively address the rough-to-smooth growth mode, advancing over qualitative comparisons with KMC simulations of recent studies \citep{toApplSS2021,empting2021,toreis2022}.
This framework creates opportunities for controlling the nanoscale morphology of thin films of $\CS$, HATCN, and other systems on weakly interacting substrates, with corresponding implications for technological applications.

\begin{acknowledgments}

The authors are grateful for the support of mutual research visits by the German Academic Exchange Service (DAAD) with funds from the German Federal Ministry of Education and Research (BMBF) and by the Brazilian agency CAPES (project 88881.700849/2022-01), all within the joint DAAD-CAPES project "MorphOrganic".   
F.D.A.A.R. acknowledges support from the Brazilian agencies CNPq (305570/2022-6) and FAPERJ (E-26/210.040/2020, E-26/201.050/2022, E-26/210.629/2023).
I.S.S.C. acknowledges support from FAPDF (00193-00001817/2023-43).
We acknowledge the European Synchrotron Radiation Facility (ESRF) for provision of synchrotron radiation facilities under proposal number SC-5641.
We thank Roody Nasro, Ainur Abukaev, Gianfranco Melis and Franziska Rapp for their help in sample preparation and characterization.

\end{acknowledgments}

\section*{Data availability}
The data collected during an experiment at the ID10-SURF beamline of European Synchrotron Radiation Source ESRF (Grenoble, France) are available at \url{https://doi.org/10.15151/ESRF-ES-1940867632} after an embargo period~\cite{dataESRF}.

\section*{Author contributions}
\label{contributions}
D.L. and I.S.S.C. contributed equally to this work.
Conceptualization, F.D.A.A.R., I.S.S.C., M.O., F.S.;
methodology, F.D.A.A.R., I.S.S.C., M.O., D.L.;
experiments, D.L., O.K., A.H., F.S.;
software, I.S.S.C., C.C.L., M.O.;
validation, I.S.S.C., C.C.L., M.O.;
formal analysis, F.D.A.A.R., I.S.S.C., M.O.;
investigation, F.D.A.A.R., I.S.S.C., C.C.L., M.O., D.L.;
resources, F.D.A.A.R., M.O.;
data curation, I.S.S.C., C.C.L., M.O., D.L., AH;
writing -- original draft preparation, F.D.A.A.R., M.O., D.L., F.S.;
writing -- review and editing, F.D.A.A.R., M.O., D.L., F.S.;
visualization, I.S.S.C., C.C.L., M.O., D.L.;
supervision, M.O.;
project administration, F.D.A.A.R., M.O.;
funding acquisition, F.D.A.A.R., M.O., F.S..

\newpage

\bibliography{interfaces}

\end{document}